\documentclass[5p,twocolumn]{elsarticle}
\usepackage{graphicx}
\usepackage{float}
\usepackage{subfig}
\usepackage{lineno}
\usepackage{amsmath}
\usepackage{fullpage}
\journal{Nuclear Instruments and Methods A}

\begin{document}

\begin{frontmatter}


\title{Event Identification in $^3$He Proportional Counters Using Risetime Discrimination}

\author[a,b]{T.~J.~Langford\corref{cor1}}
\address[a]{Department of Physics, University of Maryland, College Park MD, 20742 USA}
\address[b]{Institute for Research in Electronics and Applied Physics, University of Maryland, College Park MD, 20742 USA}
\author[c]{C.~D.~Bass\fnref{cBass}}
\address[c]{National Institute of Standards and Technology, Gaithersburg MD, 20899 USA}
\fntext[cBass]{Current Address: Thomas Jefferson National Accelerator Facility, Newport News VA, 23606 USA}
\author[a]{E.~J.~Beise}
\author[a]{H.~Breuer}
\author[a]{D.~K.~Erwin}
\author[c]{C.~R.~Heimbach}
\author[c]{J.~S.~Nico}
\cortext[cor1]{Corresponding Author: tlangfor@umd.edu} 

\begin{abstract}
We present a straightforward method for particle identification and background rejection in $^3$He proportional counters for use in neutron detection. By measuring the risetime and pulse height of the preamplifier signals, one may define a region in the risetime versus pulse height space where the events are predominately  from neutron interactions. For six proportional counters surveyed in a low-background environment, we demonstrate the ability to reject alpha-particle events with an efficiency of 99\%. By applying the same method, we also show an effective rejection of microdischarge noise events that, when passed through a shaping amplifier, are indistinguishable from physical events in the counters. The primary application of this method is in measurements where the signal-to-background for counting neutrons is very low, such as in underground laboratories. 

\end{abstract}

\begin{keyword}
Helium proportional counters \sep Neutron detectors \sep Thermal neutrons \sep Pulse shape discrimination \sep Particle identification \sep Microdischarges
\end{keyword}
\end{frontmatter}


\section{Introduction}
Helium-3 proportional counters are used as neutron detectors in a variety of fields in physics such as neutron scattering and nuclear and particle physics, particularly those experiments that operate underground in a low-background environment. They also have a commercial application as neutron detectors in health physics, medical physics, and nuclear safeguards. $^3$He has the advantages of a large thermal neutron capture cross section, easily detected charged reaction products, and low gamma-ray sensitivity. Although the basic properties of $^3$He proportional counters have been known for decades~\cite{Mills1962}, more recent applications in counting low rates of neutrons has spurred work in better understanding their backgrounds. Methods for improving background rejection in the counters~\cite{Sayres1964,Izumi1971} can be a significant benefit for those experiments or applications where the neutron signal-to-background ratio is particularly poor.

Specifically, alpha-particle decays are a pernicious background for $^3$He detectors used in low counting rate applications. The rate of alpha-particle decays originating from the walls of the counter can be comparable to, or even exceed, the rate of neutron captures. Alpha particles arise primarily from daughters in the uranium and thorium decay chains. The amount of energy deposited in the detector can be from essentially zero to several MeV, spanning the region of the neutron capture products. Thus, energy discrimination alone is not sufficient to reject background events from alpha particles. The method of risetime discrimination provides a straightforward method to reject a large fraction of these events. 

For experiments operating in deep underground environments, cosmic-ray induced neutron events can be a particularly difficult background because one cannot turn off the source of the background~\cite{Delorme1995,Formaggio2004}. The experiments can only reduce the ambient neutron background through shielding and precisely quantifying the remaining neutron flux~\cite{Mei06}. Solar neutrino experiments have used $^3$He proportional counters as triggers for neutral current interactions and monitors of muon-induced neutron backgrounds~\cite{Robertson1998}. Dark matter experiments are particularly susceptible to neutron backgrounds, due to neutron recoil events that can mimic WIMP dark matter interactions. To this end, many experiments use $^3$He proportional counters to monitor and characterize the neutron flux at their experimental site~\cite{Chazal1998,Edelweiss2010}.

In one notable example, the Sudbury Neutrino Observatory (SNO) collaboration took considerable effort in constructing ultra-low background $^3$He proportional counters containing low concentration of alpha emitters. That effort involved manufacturing proportional counters from materials with low concentrations of uranium and thorium, as well as working in a radon-free environment to prevent surface contamination~\cite{Cox2004,Amsbaugh2007}. For experiments that rely upon neutron counters as their triggers, these alpha-particle interactions in the proportional counter can be a leading source of background. Therefore, experiments such as SNO have distinguished alpha-particle background from neutron signals either through risetime analysis~\cite{Browne1999,Stonehill2005,Aharmim2011} or chi-squared fitting of the entire pulse waveforms~\cite{Martin2009}.

More generally, fast neutron spectroscopy can be done directly using a $^3$He proportional counter~\cite{Batchelor1955,Chichester2012}, and it is also routinely done using Bonner spheres, where fast neutrons are moderated in a hydrogenous material and a fraction of those thermalized neutrons may be captured by a central $^3$He proportional counter~\cite{Thomas2002,Wiegel2002}. Another technique for fast neutron detection using $^3$He proportional counters is capture-gated spectroscopy~\cite{Abdurashitov2002}. The fast neutrons recoil from protons in an active hydrogenous moderator; the moderated neutron diffuses in the medium and may be subsequently captured on $^3$He.  The time-delayed capture signifies that the preceding interaction was a fully thermalized neutron. This allows for an accurate measurement of the incident particlesÕ energy, as well as a highly efficient method for background reduction.

Work in nuclear safeguards has made helium-based neutron detectors the focus of detecting special nuclear material. Most neutron-detecting portal monitors use a $^3$He proportional counter surrounded by a passive neutron moderator~\cite{Kouzes2010}. Because the monitors are searching for very low levels of neutron activity, the technical requirements for these detectors and those operating in an underground environment are very similar. In this case, background events are false positives, and hence alpha particles and noise in the detectors are a detriment to the effectiveness of the portal monitors~\cite{Kouzes2011}.

Improving the particle identification and rejection of events in $^3$He proportional counters that are not neutrons has clear relevance to these applications. More specifically, this collaboration is constructing a fast neutron spectrometer~\cite{Langford2010} that employs a large number of $^3$He counters, and hence it is critical to understand all the background contributions in each counter that may be part of the spectrometer. Approximately 75 $^3$He proportional counters of the same model were surveyed as part of this study.

In this paper, we discuss how a combination of risetime and energy analysis can significantly improve the identification of neutron capture events in a $^3$He proportional counter. Section~\ref{sec:HeOperation} briefly reviews the operation of $^3$He proportional counters. In Section~\ref{sec:Setup} we describe the experimental setup and analysis method for testing $^3$He proportional counters. Section~\ref{sec:ParticleID} gives the approach for identifying the origin of various waveforms that may occur in a $^3$He proportional counters. We apply that method to quantify the rejection of non-neutron events for a specific counter in Section~\ref{sec:bkgd_rej}. Section~\ref{sec:SBD} describes an auxiliary measurement to quantify alpha-particle rates from the proportional counter body, and Section~\ref{sec:concl} summarizes the results. 

\section{$^3$He proportional counters}
\label{sec:HeOperation}
\subsection{Principles of operation}

$^3$He provides an excellent medium for detecting thermal neutrons due to the high capture cross section ($5330\times 10^{-24}$\,cm$^2$ ) and a final state that consists of charged particles, as opposed to gamma rays. A neutron is captured by a $^3$He nucleus, resulting in a proton and a triton, which share 764\,keV of kinetic energy

\begin{equation}
n \; + \; ^3\textrm{He} \rightarrow p \;+\; t\; +\; 764 \textrm{\;keV.}
\end{equation}
These heavy charged particles ionize the gas in the proportional counter, creating electron-ion pairs that are proportional to the energy deposited. For cylindrical proportional counters, the electrons drift towards a central anode wire, and when the electric field gradient reaches a critical level, begin to avalanche. Thus, the detected current is created from the drifted charge.

Because the neutron capture on $^3$He has a two body final state, the charged particles will be emitted in opposite directions. If the ionization tracks are parallel to the central anode wire, all of the charge will be collected at approximately the same time, leading to a short risetime of the detected signal. However, if the particles are emitted perpendicular to the anode wire, one particle will be moving towards the wire while the other is moving away.  The collected charge will be spread out in time due to the radial variation caused by the track geometry~\cite{Takeda1995}. 

Typically, a $^3$He proportional counter is biased through a preamplifier, and the resulting signals are sent through a shaping amplifier to be analyzed with a multi-channel analyzer or peak-sensing analog to digital converter. However, shaping the preamplifier signal smoothes out the slight deviations in signal shape caused by the track geometry. With high-speed waveform digitizers, it is possible to study the shape of preamplifier signals directly and apply digital signal processing to identify the type of incident radiation. 

\subsection{Specific energy loss and particle range}
\label{sec:Theory}
The specific energy loss of charged particles with charge $Z$, mass $m$, and kinetic energy $E_{kin}$, is approximately proportional to

\begin{equation}
-\frac{dE}{dx} \propto \frac{Z^2 m}{E_{kin}}.
\end{equation}

Thus, the track length for a given $E_{kin}$ decreases with increasing mass and charge of the projectile. For neutron capture on $^3$He, the proton is emitted with 573\,keV and the triton with 191\,keV. At these energies and using the method of Anderson et al.~\cite{Andersen1977}, the proton has a range of  1.88\,mg/cm$^2$ and the triton has a range of 0.71\,mg/cm$^2$ in our proportional counters. Because the proton and triton are back-to-back, this yields a total track length of 2.6\,mg/cm$^2$, compared to a 764\,keV alpha particle, for example, with a range of 1.0\,mg/cm$^2$. 

Beta emitters, electrons from $\gamma$-interactions, and cosmic rays leave long tracks and deposit little energy resulting in small signals with a wider range of risetimes. Alpha particles with much higher specific energy loss leave shorter tracks and deposit more of their energy, which yield large signals with a relatively fast risetime. These features can be exploited in analysis to help identify the original radiation. Details about the energy loss of charged particles and the resulting pulse shapes in proportional counters can be found elsewhere~\cite{Andersen1977,Leo1994,Takeda1995,Knoll2000,Beltran2011}.

\section{Experimental setup}
\label{sec:Setup}

\begin{figure}
\begin{center}
\includegraphics[scale=.4]{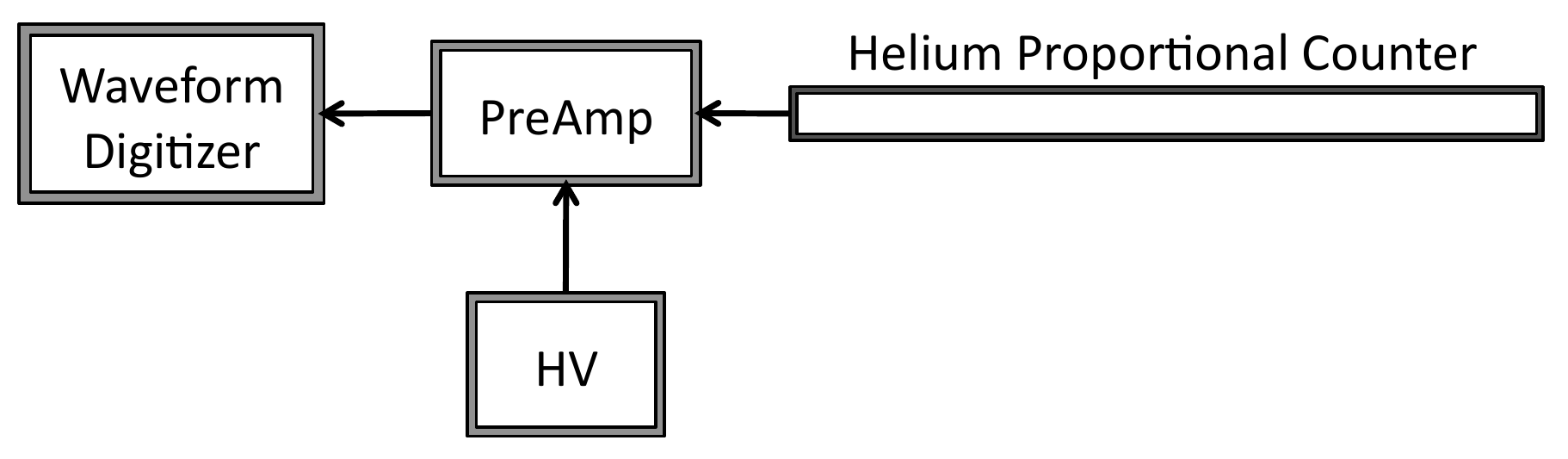}
\caption{A schematic showing the setup for a single channel of our measurements. Particles in the $^3$He proportional counters generate current signals that go to the preamplifier. The waveform digitizer captures the resulting long-tailed signals and records them for off-line analysis.}
\label{fig:SetupDiagram}
\end{center}
\end{figure}

The experimental setup is shown in Fig.~\ref{fig:SetupDiagram}. We used 2.54\,cm outer diameter, 46.3\,cm active length, aluminum-bodied, cylindrical $^3$He proportional counters manufactured by GE-Reuter Stokes\footnote{Certain trade names and company products are mentioned in the text or identified in an illustration in order to adequately specify the experimental procedure and equipment used.  In no case does such identification imply recommendation or endorsement by the National Institute of Standards and Technology, nor does it imply that the products are necessarily the best available for the purpose.}. The $^3$He partial pressure in the counters was 404\,kPa (4\,atm) with a buffer gas consisting of 111\,kPa (1.1\,atm) of krypton. Helium does not have a high stopping power for heavy charged particles. Therefore, manufacturers tend to add in a buffer gas with higher stopping power to ensure full energy deposition from the capture products. The krypton increases the stopping power of the gas for charged particles but has little effect on neutron capture. 

\begin{figure}
\begin{center}
\includegraphics[scale=.27]{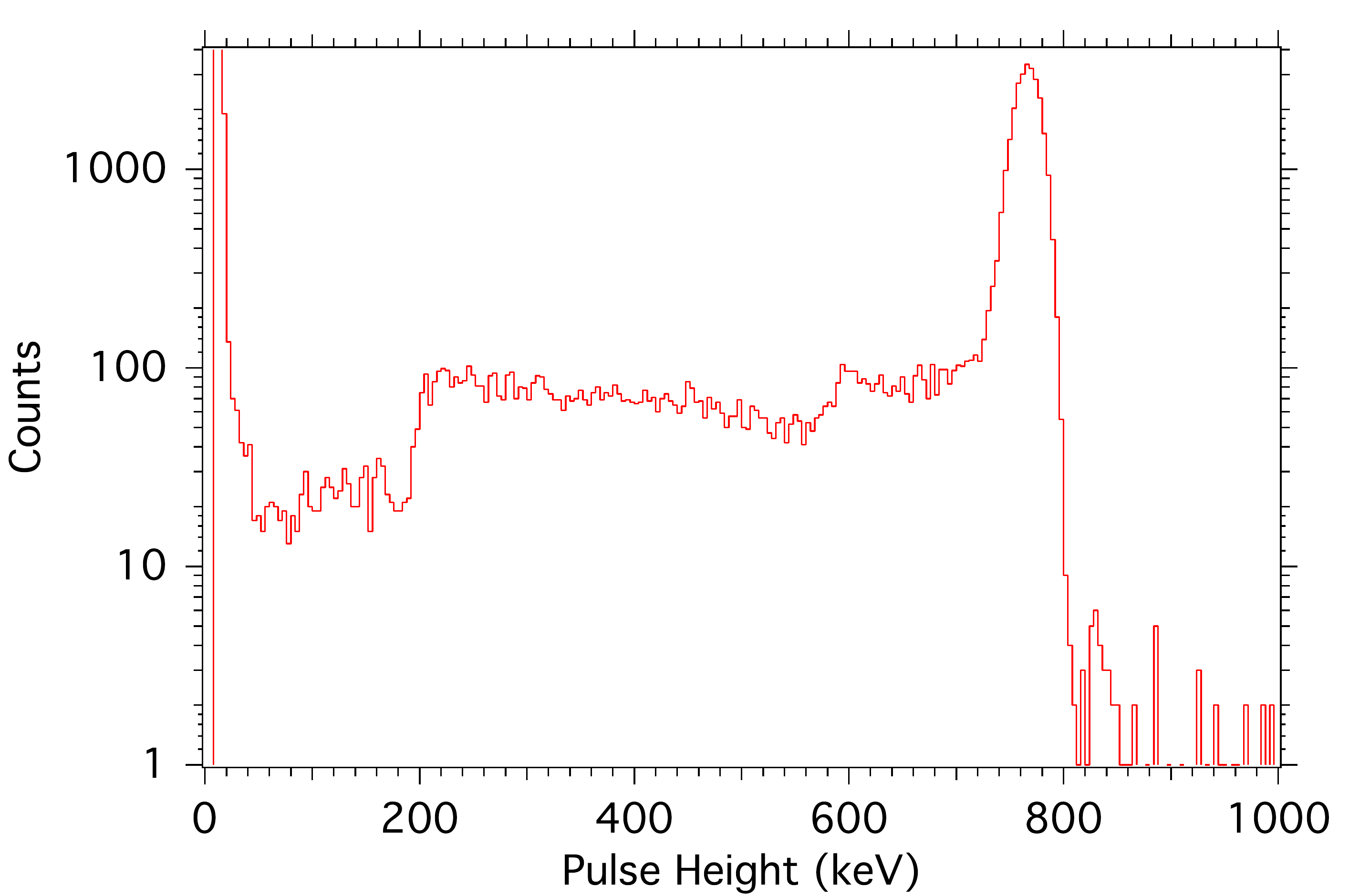}
\caption{Typical energy spectrum of neutron capture events in one of the $^3$He proportional counters used in this study. The dominant feature is the full-deposition peak at 764\,keV. The two edges in the spectrum near 200\,keV and 600\,keV are related to partial energy deposition in the gas
when either the proton or the triton interacts in the wall.}
\label{fig:energySpec}
\end{center}
\end{figure}

The proportional counters were biased through a charge sensitive preamplifier, and the traces were recorded for offline analysis by a 125\,MSamples/s 12-bit GaGe waveform digitizer or a 250\,MSamples/s 12-bit CAEN waveform digitizer. Data were taken at different voltage levels to test the variability of the counters' response. For these specific counters, 2000\,V was  a reasonable operating voltage. An exploration of the effect of different counter bias voltage indicated that there was little variation of the risetime distribution.   

Figure~\ref{fig:energySpec} shows a typical energy spectrum measured by one of the proportional counters used in this work. Note that although the neutron capture reaction is monoenergetic, there are still features in the energy spectrum related to when one of the particles interacts in the counter wall, leading to reduced energy deposition. This well-known wall effect is clearly seen in the two edges around 200\,keV and 600\,keV in Fig.~\ref{fig:energySpec}. Events below 200\,keV are largely betas liberated from the counter body by gammas from the $^{252}$Cf source.

\subsection{Analysis method}

\begin{figure}
   \centering
   \includegraphics[scale=0.27]{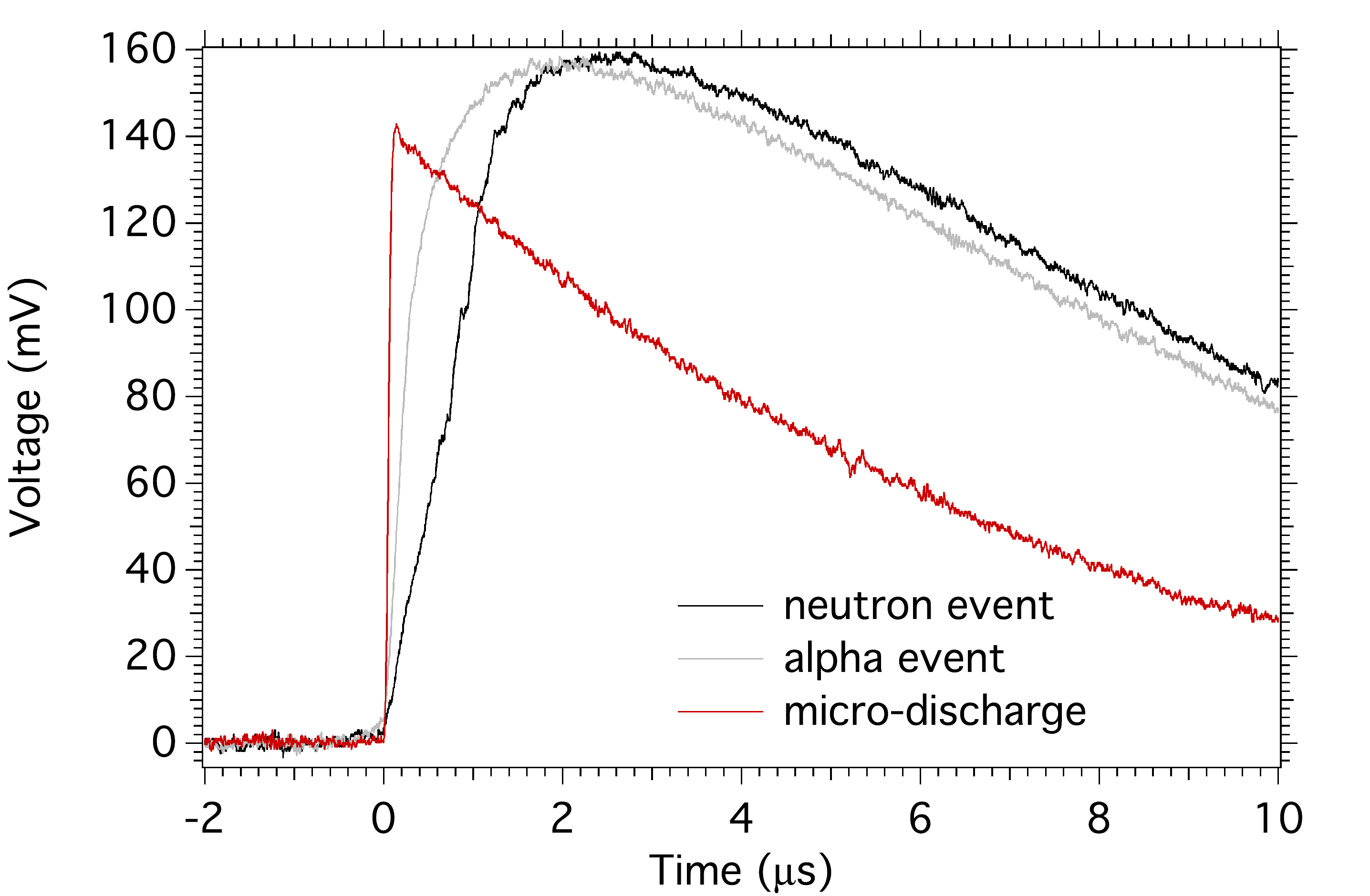} 
   \caption{Example of the raw preamplifier traces showing the risetimes from three different sources. Shown here, in sequence of increasing risetime, are a microdischarge (red), an alpha-particle trace (gray) and a neutron trace (black).}
   \label{fig:signalEx}
\end{figure}

The digitized waveforms were analyzed, and both the amplitude and risetime of the signals were extracted. Figure~\ref{fig:signalEx} shows sample traces corresponding to three different sources: a neutron event, an alpha-particle event, and a microdischarge from the proportional counter. Microdischarges are electronic artifacts originating from current leakage in the bias voltage breakdown applied to the counter~\cite{Kitani1994,Danikas1997a,Danikas1997b}. To determine the pulse height of the waveform, one first removes the baseline offset by subtracting off the average of the first few hundred samples of the trace, well before the pulse onset. To remove high-frequency noise, a Gaussian smoothing routine was applied to the waveform, and the amplitude was determined from the resulting wave. Because the signals were sent though a charge-sensitive preamplifier, the energy of the particle is proportional to the amplitude of the recorded signal. The fall time of the signals were a property of the AC coupling and decay time of the preamplifier. The detectors were calibrated using the thermal neutron capture peak to set the energy scale.

For this work, the risetime was defined as the time difference between the signal reaching 10\% and 50\% of its maximum amplitude. This is the same interval used in Ref.~\cite{Browne1999}. This choice was based upon a reasonable separation of the three signal types. If the interval was too small (less than 20\%), there was not adequate separation between the microdischarges and the real signals. However, the interval could not be made too large, otherwise the slow component of the risetime, due to the drift of positive ions~\cite{Knoll2000}, decreases the separation of alpha-particle and neutron events. For both the alpha particles and neutron traces shown in Fig.~\ref{fig:signalEx}, this begins at about 75\% of the maximum signal amplitude.

\section{Event identification through risetime analysis}
\label{sec:ParticleID} 
\subsection{Physical signals from source radiation}

\begin{figure*}
\begin{center}
  \subfloat{\label{fig:nScatter}}\includegraphics[scale=0.27]{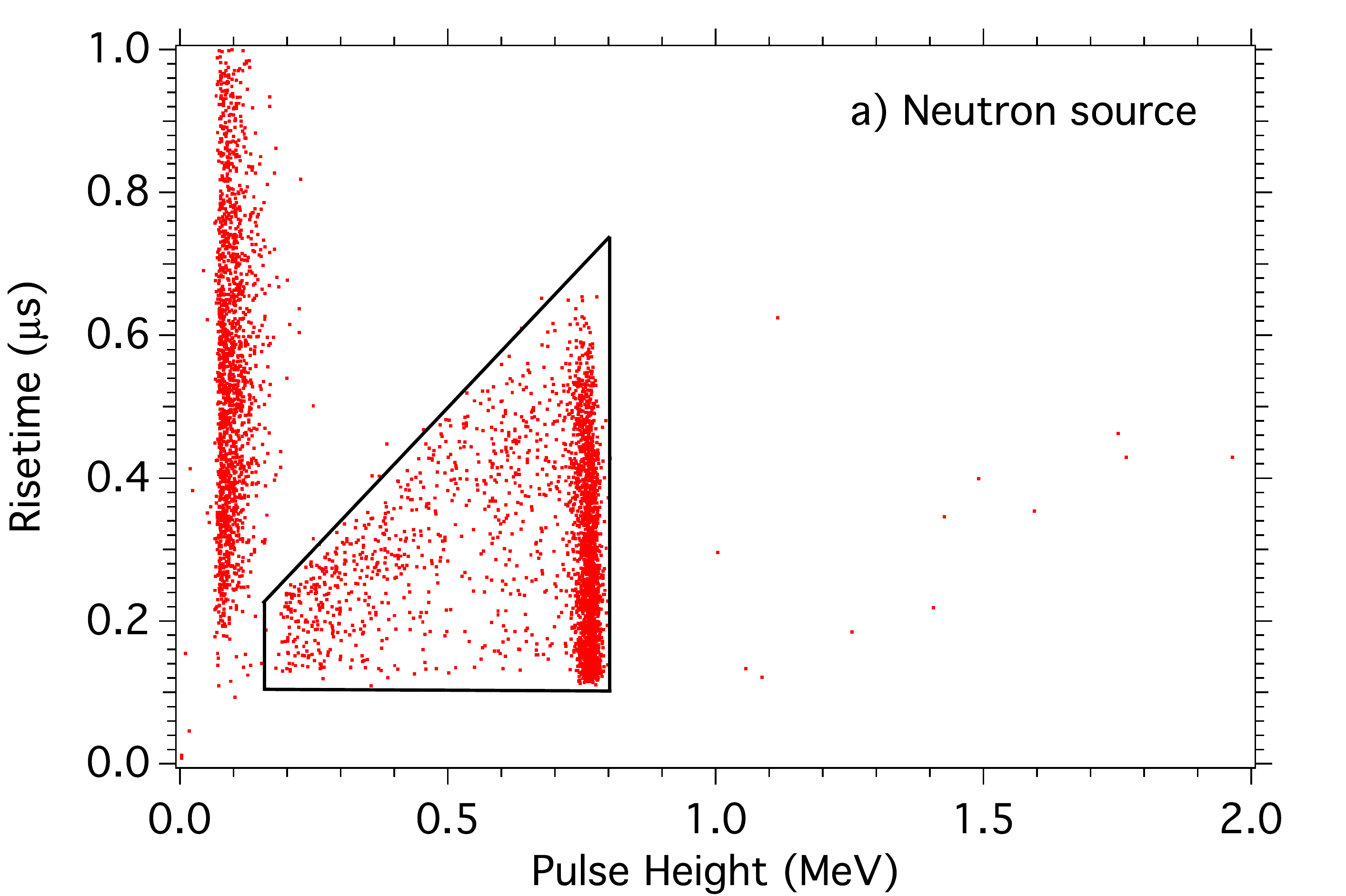}
  \subfloat{\label{fig:gScatter}}\includegraphics[scale=0.27]{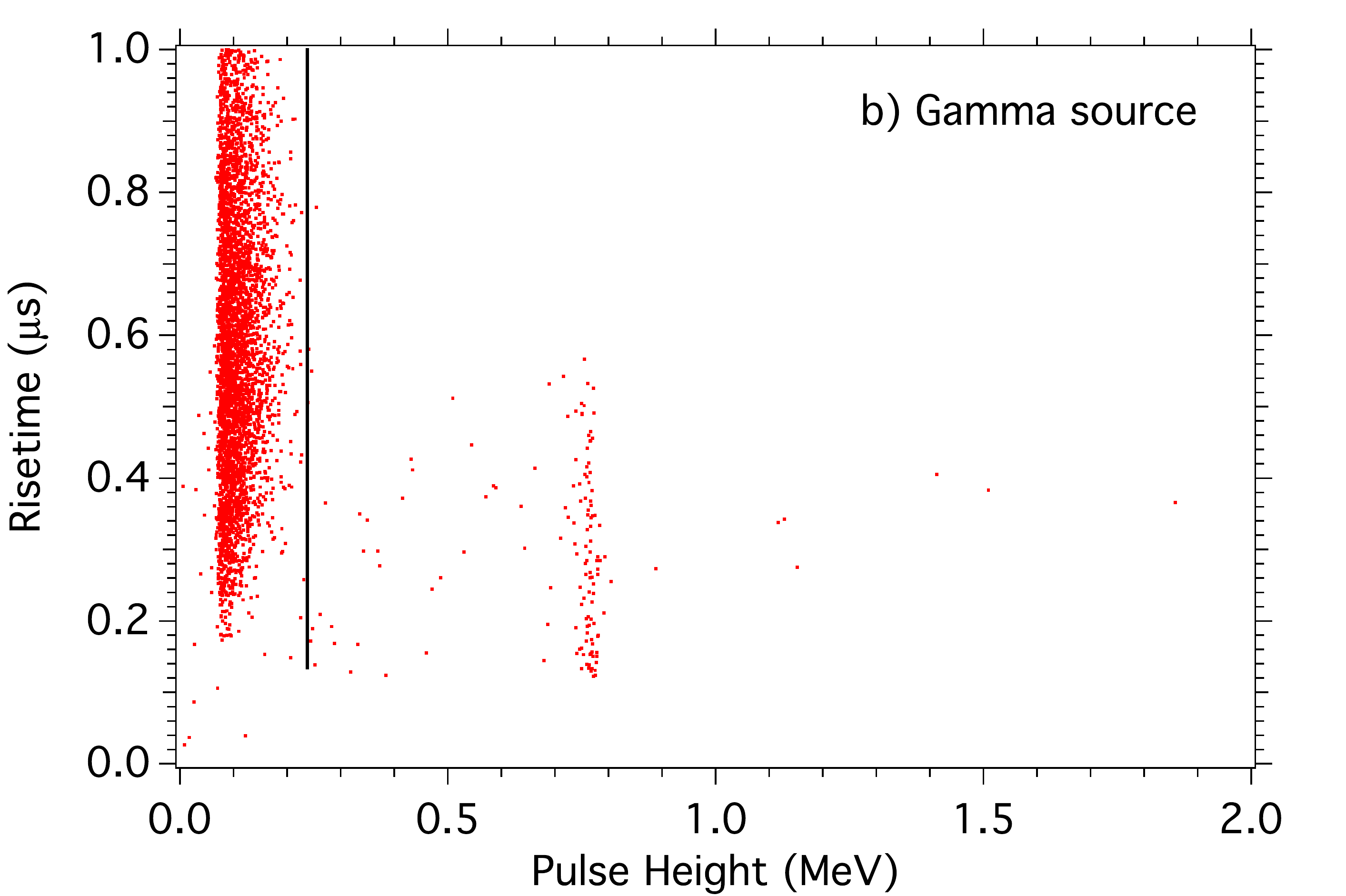}              
  \subfloat{\label{fig:bScatter}}\includegraphics[scale=0.27]{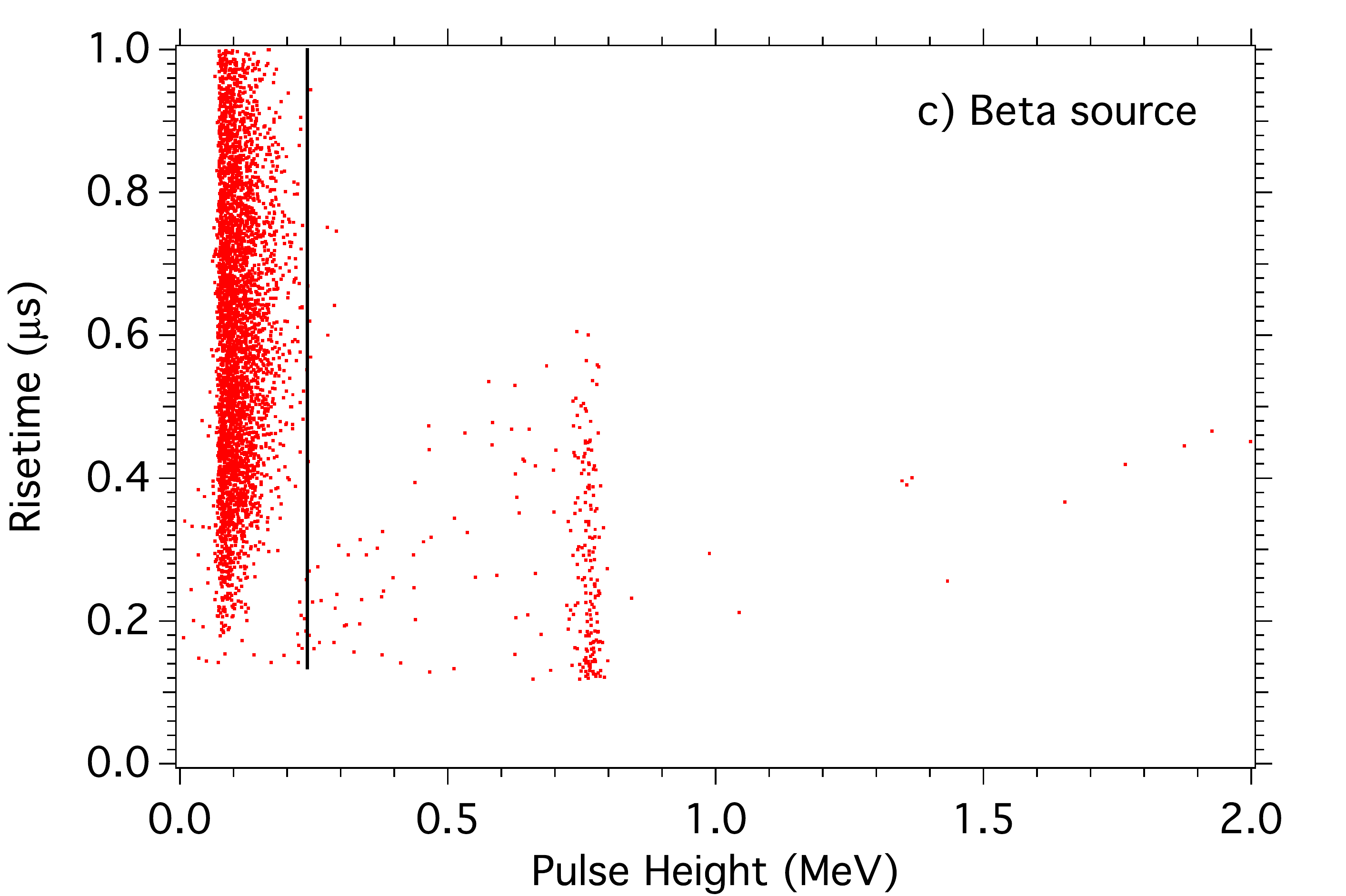}
  \subfloat{\label{fig:aScatter}}\includegraphics[scale=0.27]{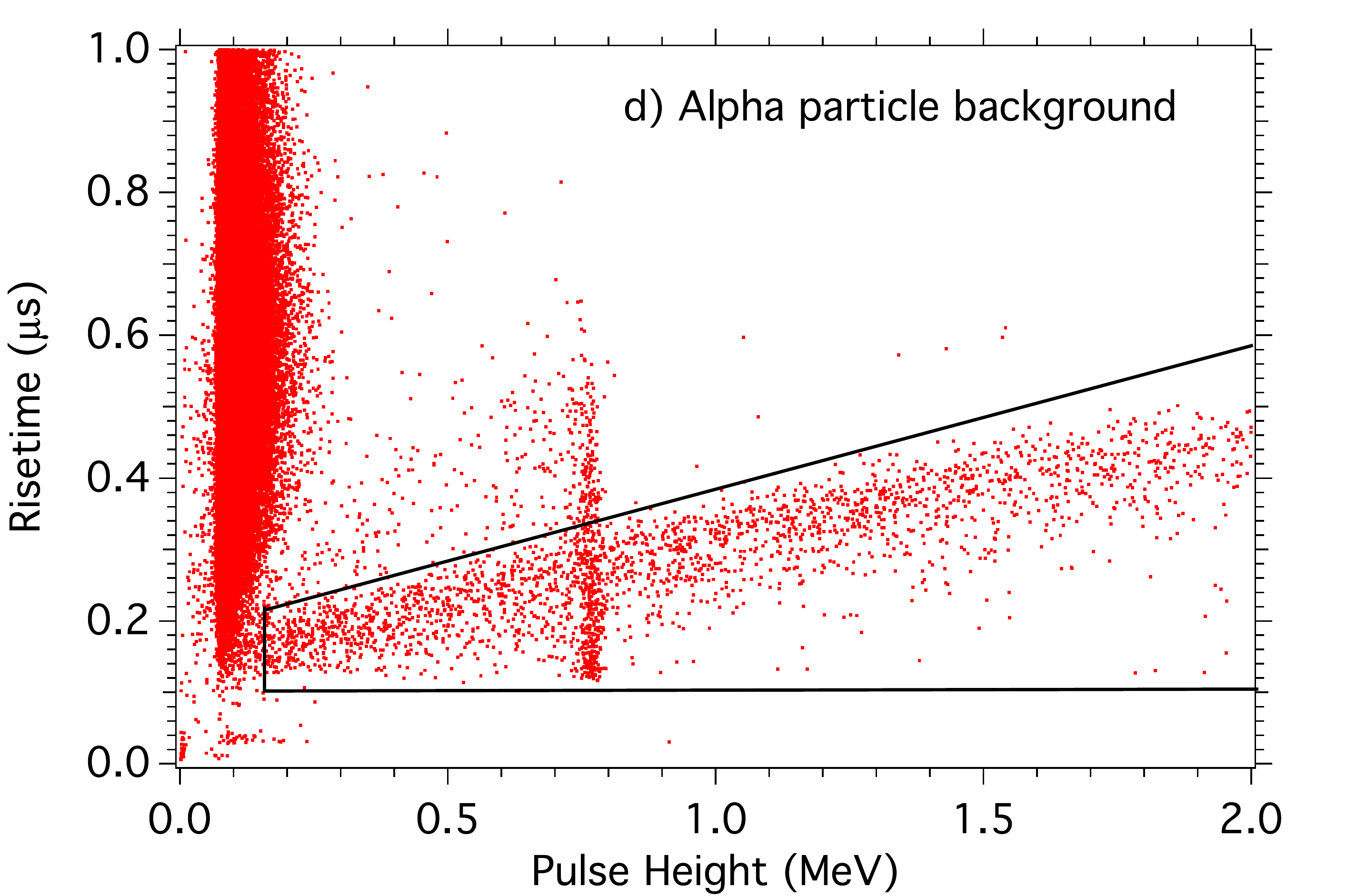}
  \subfloat{\label{fig:mdScatter}}\includegraphics[scale=0.27]{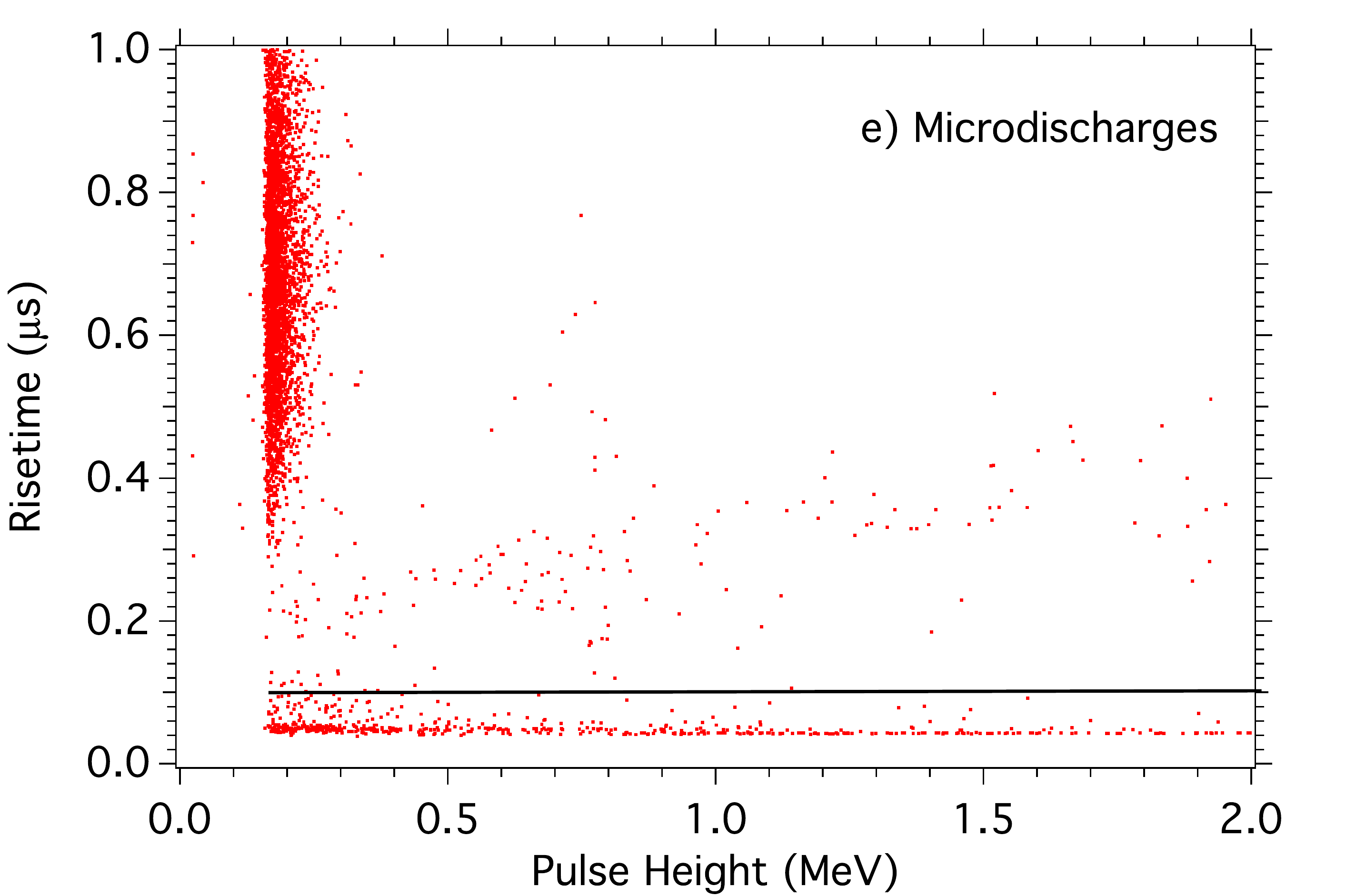}
\caption{Scatter plots showing the risetime versus pulse height from different sources of radiation and microdischarges in $^3$He proportional counters:  a) neutron source data; b) gamma source data; c) beta source data; d) alpha-particle background data; and e) microdischarges. The solid black lines illustrate regions where the indicated events occur. Further description is found in the text of Section~\ref{sec:ParticleID}.}
\label{fig:sourceScatter}
\end{center}
\end{figure*}

Events in the counters arise from other radiation sources or bias voltage breakdown in the counter. Data were collected under a few different conditions to characterize the $^3$He proportional counters' response to different sources of radiation. By displaying the risetime versus the pulse height of each signal, it is possible to identify regions of this two-dimensional space where events from different source radiation will occur. Figure~\ref{fig:sourceScatter} shows the proportional counter response to thermal neutron, gamma, beta, alpha, and microdischarge events. 

Thermal neutrons were generated from neutrons from $^{252}$Cf that were moderated by a 5\,cm thick piece of polyethylene placed between the source and proportional counter. Figure~\ref{fig:sourceScatter}a illustrates the region in both risetime and pulse height where the counter detects the neutron-capture reaction products. The large concentration of events around 764\,keV correspond to the full energy deposition. The risetime varies depending upon the orientation of the decay products in the counter with respect to the central anode wire.

To test the counter's response to beta and gamma radiation, $^{90}$Sr and $^{60}$Co sources were placed directly above the counter in separate measurements. Figures~\ref{fig:sourceScatter}b and \ref{fig:sourceScatter}c show that these interactions are comparatively low energy that can be easily discriminated from neutron events. Note that there are still neutron events from the ambient background in the laboratory where the measurements were performed.

Figure~\ref{fig:sourceScatter}d illustrates why alpha particles are the most difficult background to address. Many events  overlap with neutrons in both energy and risetime. For this measurement, the source of the alpha particles was the trace amount of uranium and thorium contamination in the wall of the proportional counter body.  To emphasize the alpha-particle events, the counter was wrapped in boron-loaded silicone rubber to absorb most of the external ambient background of thermal neutrons.  As the typical alpha-particle detection rates for the counters in this work are quite low, data were taken over the course of a few days to acquire sufficient statistics. Note that with such a long run, there is still some leakage of thermal neutron events into the spectrum.

We characterized the events from about 75 $^3$He proportional counters for this risetime study. We note that they had a wide range of alpha activity as well as variations in spectral shape, even though they all came from the same manufacturer. Some counters had rates consistent with or higher than the raw aluminum expectation, discussed in Section~\ref{sec:SBD}, while some were considerably lower. The counter with the lowest alpha-particle rate had approximately $6 \times 10^{-4}$\,s$^{-1}$, and the highest alpha-particle rate counter was approximately $3 \times 10^{-2}$\,s$^{-1}$. Certain counters exhibited a broad spectrum of alphas with no peaks while others displayed peaks at energies consistent with isotopes in the decay chains of uranium and thorium. During construction of the counters, a thin layer of nickel was deposited onto the inner surface of the cylinders. This was done by the manufacturer to minimize the contribution of alpha particles to the background~\cite{johnsonEmail}. We speculate that deviations in the thickness of this coating could cause the observed variation.

\begin{figure}
\begin{center}
\includegraphics[scale=0.27]{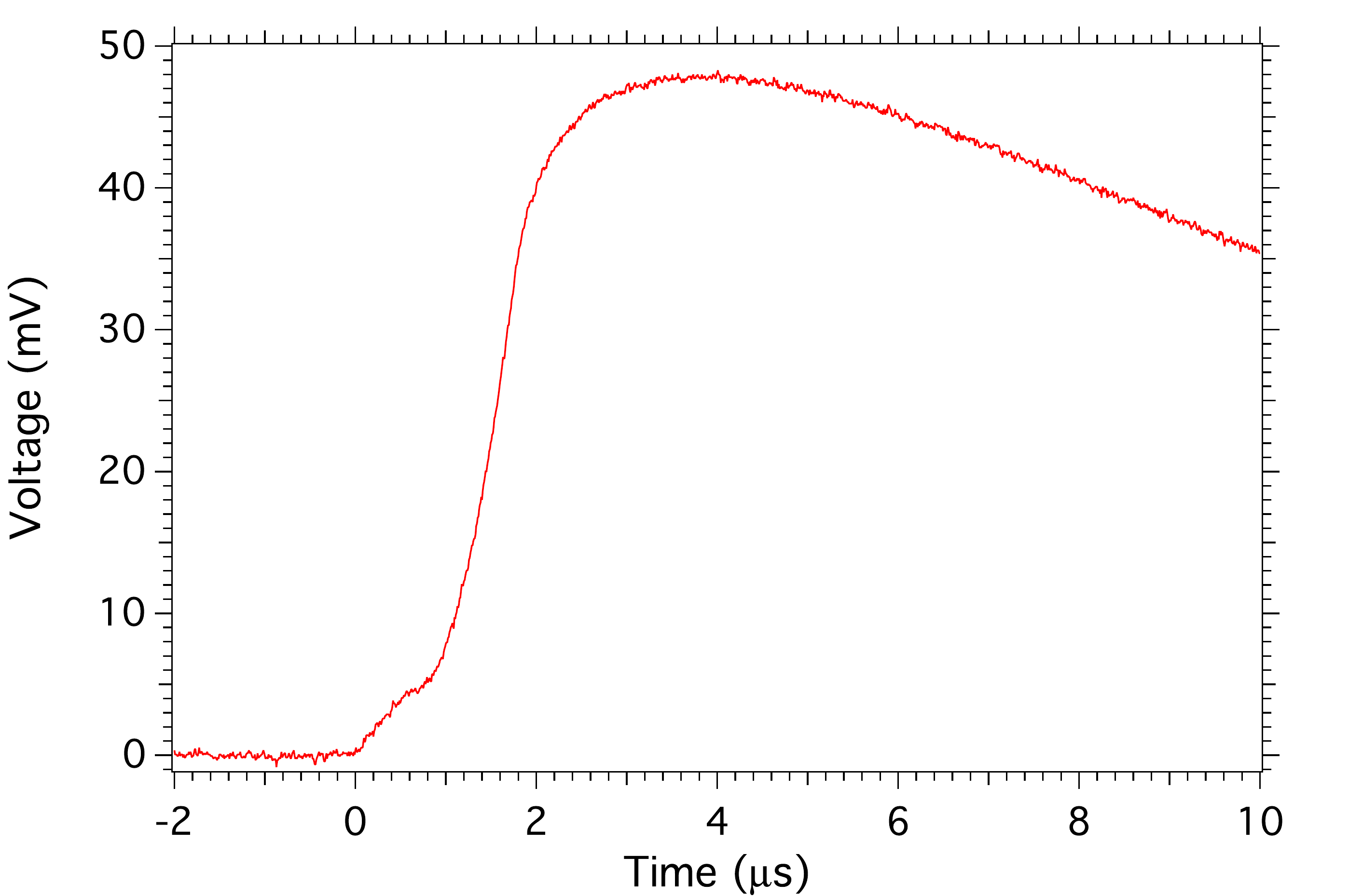}
\caption{A sample trace attributed to an alpha-particle and gamma-ray event in the proportional counter. These are possibly from a correlated decay, where a nucleus alpha decays to an excited state, which then de-excites by emitting a gamma.}
\label{fig:alphaGammaPileup}
\end{center}
\end{figure}

In counters with high alpha-particle rates, events were observed with pileup of a large alpha-like signal on top of a small gamma-like signal.  These events are possibly from correlated decays, where a nucleus alpha-decays to an excited state, which subsequently emits a gamma. An example trace of this process is shown in Fig.~\ref{fig:alphaGammaPileup}. The low-amplitude gamma signal will appear first because its track is extended and closer the anode. The alpha particles predominately originate from the counter walls, the furthest distance from the anode, and therefore arrive after the gamma event. As a result, these signals have longer risetimes, which could lead to misidentification as neutron-like signals with this analysis. Other techniques, such as waveform fitting, should be able to identify such events.

\subsection{Microdischarges}

\begin{figure*}
\begin{center}
\includegraphics[width=.9\textwidth]{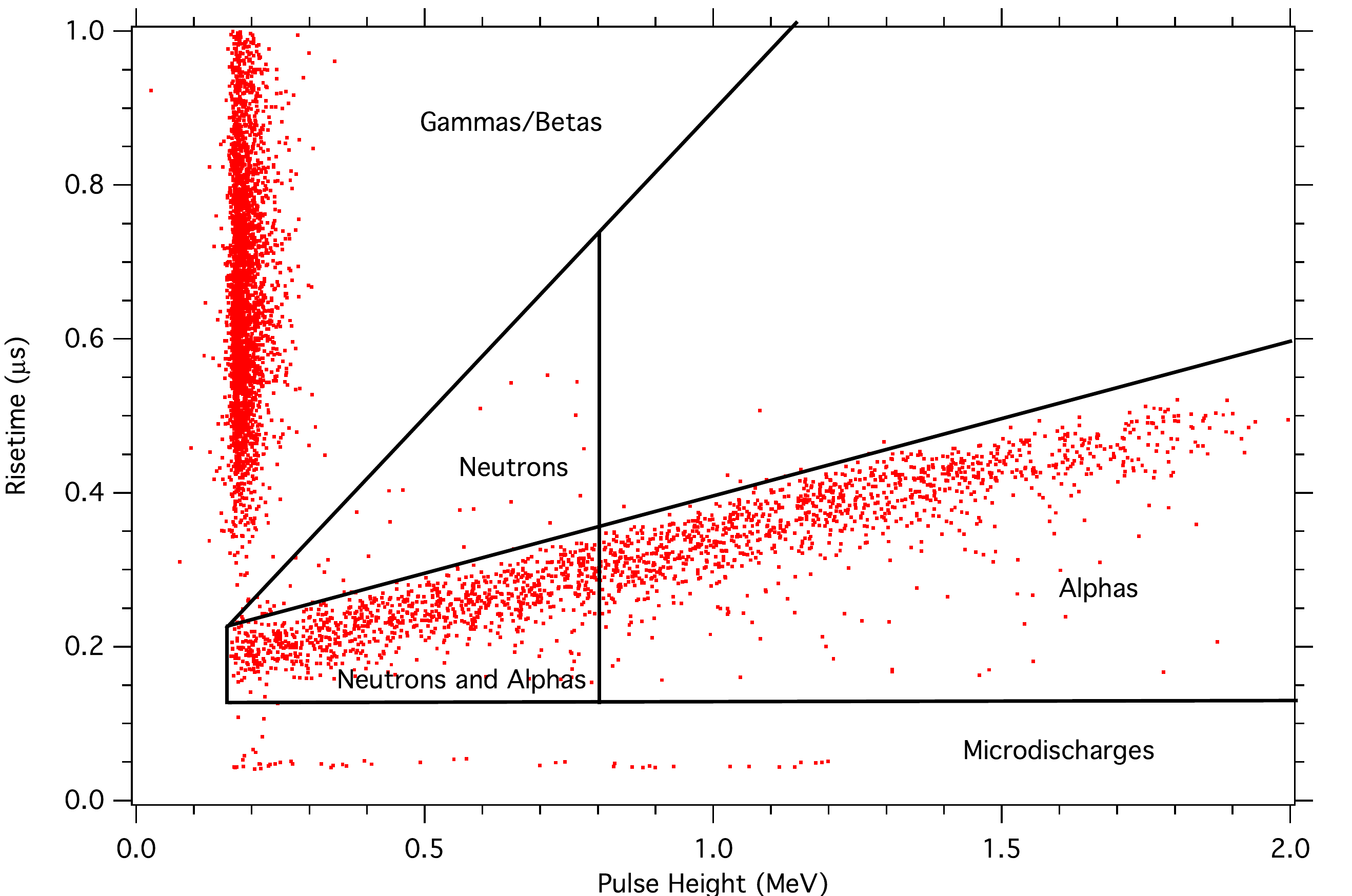}
\caption{Risetime versus pulse height data from a  $^3$He proportional counter placed underground at KURF. The black lines and labels serve to indicate the predominate type of event in the region. The low neutron flux facilitates the study of the internal alpha-particle contamination of the counters. This counter has a broad alpha spectrum and a low microdischarge rate, as shown in the labeled regions.}
\label{fig:cutScatter}
\end{center}
\end{figure*}

In the course of the measurements, a few counters were found to have spurious signals with very rapid risetimes. After studying the counters in question, the source was determined to be microdischarges from the high voltage feedthrough to the grounded case of the counter. An example of such a signal is shown in Fig.~\ref{fig:signalEx} along with neutron-capture and alpha-particle signals. This is a known effect in $^3$He proportional counters, and a thorough treatment of the origin of these signals can be found in Ref.~\cite{Heeger2000}. The microdischarge is seen by the preamp as a current pulse and is treated as a normal signal. When sent through a shaping amplifier, there is no way to distinguish a discharge from a signal generated by an incident particle. By using the preamplifier signal, it is possible to measure the fast risetime of these spurious signals and discriminate against them.   Figure~\ref{fig:mdScatter} shows an example of a detector with a clear band of fast ($<100$\,ns) risetime events that span a wide range of pulse height. The rate of such events may vary significantly among proportional counters, even for counters made by the same manufacturer.

\section{Non-neutron event rejection using energy and risetime discrimination }
\label{sec:bkgd_rej}

Utilizing the energy and risetime of these signals provides a straightforward and fast method to characterize the shape and origin of events in a counter, as well as a rationale to define cuts that minimize backgrounds specific to a given application. Figure~\ref{fig:cutScatter} is a risetime versus pulse height plot from one of the $^3$He proportional counters placed underground.  The black lines indicate where one may apply cuts in an analysis to isolate the different events.  Two methods of performing cuts on the data proved useful in this work and are described in this section. First, one may place restrictive cuts that define an essentially neutron-only region,  This is particularly important for the applications where the rate from the neutron signal is expected to be very low.  Second, one may place less restrictive cuts that eliminate microdischarges and gamma and beta backgrounds but accept the contribution from alpha-particle contamination. 

\subsection{Neutron-only cut region}

In the risetime versus pulse height space, one can define a region essentially free of alpha and beta/gamma backgrounds, as well as microdischarge noise.  This region is clearly shown in Fig.~\ref{fig:cutScatter}. Three cuts are required to isolate neutron events: two diagonal cuts and one in pulse height only. The first diagonal cut is defined by events above the upper edge of the alpha region. The second diagonal cut is defined as events below the upper edge of the distribution from neutron source data. Finally, an upper energy threshold is made to reject events with energy above 0.8 MeV, which is the approximate end of the neutron capture peak. These cuts define a triangular region which is essentially free from alpha-particle and beta/gamma backgrounds and microdischarge events.

To test leakage of alpha-particle events into the neutron-only region, we operated several of the proportional counters in a low-neutron environment at the Kimballton Underground Research Facility (KURF) in Blacksburg, VA. The facility is located in an active limestone mine at a depth of 1450 meters water equivalent and provides a good low-radioactivity counting environment~\cite{Finnerty2010}. Six helium detectors were counted for approximately 16 hours. These detectors showed a variety of alpha and microdischarge rates;  the risetime versus pulse height spectra from one of the counters is shown in Fig.~\ref{fig:cutScatter}. 

It is possible to estimate the leakage of alpha events into the signal region by looking at events above the neutron capture peak in Fig.~\ref{fig:cutScatter}. There should be no neutrons in this region, but the alpha events should have the same probability to spill into the cut-region. For the six detectors tested at KURF, we see an average of 1\,\% of alpha events which are not cut by this method.  When these cuts are applied to neutron source data, we find that approximately 55\,\% of the neutrons are rejected. The location of the cuts can be refined to match distributions present in specific counters, as well as the desired signal-to-background ratio.

\subsection{Elimination of microdischarge noise only}

After surveying the 75 proportional counters using the risetime method, we selected counters that contained the lowest alpha-particle backgrounds to minimize the contamination of the neutron signal region. This permits one to focus on the microdischarge backgrounds.  Because the microdischarge region is well-separated from the neutron region, we can use this cut to eliminate the microdischarge events while preserving the entire neutron region.

Without utilizing the information in the pulse risetime, there is nothing to distinguish microdischarge events from alpha-particle events. Thus, more proportional counters would have been categorized as high-background that are low in alpha contamination but high in microdischarges. Counters that would otherwise have been unsuitable for low background experiments can still be used, as long as the microdischarge events are rejected. 

\section{Direct measurement of alpha-particle rates from $^3$He counter walls}
\label{sec:SBD}

To verify that the background attributed to alpha particles was being correctly identified, we performed a separate measurement of the alpha activity at the surface of a $^3$He proportional counter using a silicon detector. These measured rates were compared with existing measurements~\cite{HashemiNezhad1998,AlBataina1987} and the internal alpha rates from the counters themselves. The comparison will not exact because some parameters among the counters cannot be controlled, such as the manufacturing process and knowledge of an absorbing layer deposited on the inner surface.

\begin{figure}
   \centering
   \includegraphics[scale=0.27]{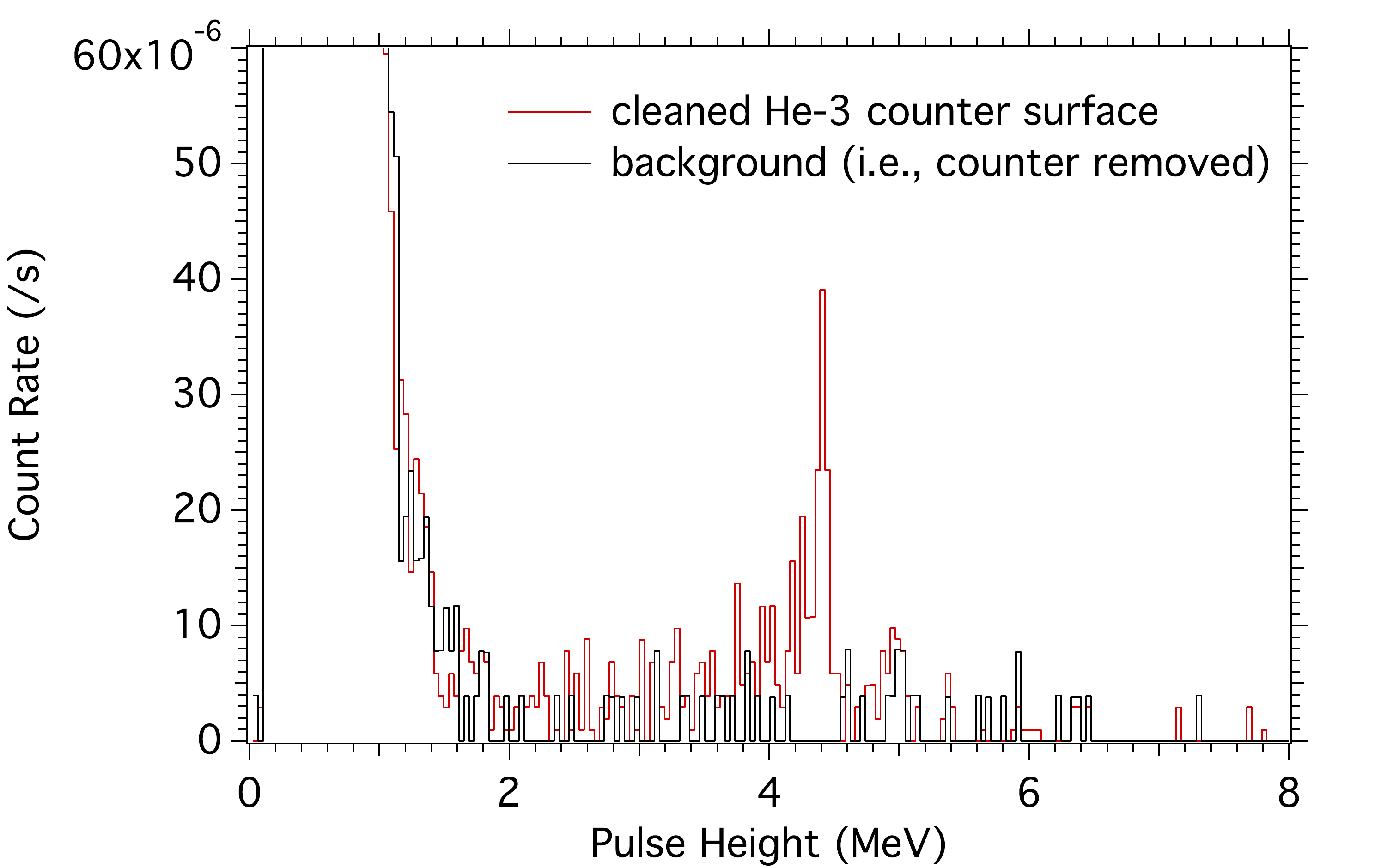} 
   \caption{Energy spectra recorded with a silicon surface barrier detector. The red trace shows the spectrum from the outside surface of a $^3$He proportional counter that had been scrubbed clean to remove surface contamination. The blue trace shows a background spectrum obtained when the $^3$He counter was removed. The broad spectrum of alpha-particle energies is consistent with bulk contamination of the counter body material.}
   \label{fig:SBDspectrum}
\end{figure}

To measure the alpha-particle rate from the aluminum surface, a $^3$He proportional counter was placed inside a vacuum chamber, and a silicon surface barrier detector was placed on the surface. The chamber was lined with thin plastic and evacuated to minimize events originating from the chamber and radon in the atmosphere, respectively. The silicon detector signal was processed through a preamp and shaping amplifier, and the peak heights were registered on a multichannel analyzer. Data were collected under a number of different conditions in order to identify sources of background. It is important to differentiate between alphas that originate from the metallic surface of the $^3$He counter and those that may come from the chamber, detector mounting hardware, or the silicon detector itself. Contamination on the surface the $^3$He counter body is a particular concern, and care was taken to clean the surface with a mild abrasive and detergent before measurements were performed.

Figure~\ref{fig:SBDspectrum} shows the energy spectrum of particles from a $^3$He counter along with a background spectrum when the counter was removed. To compare with values in the literature, we summed the counts between 2\,MeV and 10\,MeV after background subtraction when the counter was removed. Taking into account the detector solid angle and scaling up to the full size of the proportional counter, a typical alpha rate was $3.9 \times 10^{-2}$ s$^{-1}$. Using the internal surface area of  370\,cm$^2$ for a counter, one obtains a count rate per area of $ 1.1 \times 10^{-4}$ cm$^{-2}$s$^{-1}$. In Ref.~\cite{AlBataina1987}, researchers tabulated alpha count rates per area ranging from $1.4 \times 10^{-5}$ to $1.2 \times 10^{-4}$ cm$^{-2}$s$^{-1}$ from commercial aluminum. Rates will vary widely due to several factors, such as the supplier and surface contamination~\cite{Debicki2011}, but the values obtained with this proportional counter are consistent with both the literature and our own internal measurements. This gives additional confidence that the alpha particle in the proportional counters are being correctly identified.

\section{Summary}
\label{sec:concl}

We have detailed a straightforward method to identify neutron, alpha, gamma and beta, and microdischarge events from $^3$He proportional counter preamplifier signals collected with waveform digitizers. By using the risetime and energy information of each signal, one can isolate neutron capture events and eliminate contamination of the neutron signal from the gamma and beta backgrounds and microdischarges. We have verified that the dominant background in the neutron risetime and energy window arises from alpha decays from the counter body itself. We have identified a set of selection criteria to improve significantly the neutron signal-to-background ratio in that region. If counting statistics are sufficient for given experiment, or if one desires the highest signal-to-background ratio, one may choose to sacrifice approximately half of the neutron events in order to eliminate nearly all of the alpha-particle events. In many low event rate applications, that may not be the preferred option; in which case, this method will provide an important tool to identify and mitigate the contribution of alpha backgrounds in $^3$He proportional counters. 

A fast neutron spectrometer utilizing six $^3$He proportional counters~\cite{Langford2010} was placed underground at KURF. Using the risetime method to identify and reduce backgrounds has led to a significant improvement of the experiment. In future work, we will investigate the improvement in the neutron signal-to-background from data acquired in a low-background environment. 

\section{Acknowledgments} 
We thank Vladimir Gavrin and Johnrid Abdurashitov of the Institute for Nuclear Research - Russian Academy of Sciences for useful discussions. We acknowledge the NIST Center for Neutron Research for the loan of the $^3$He proportional counters used in this study. We also acknowledge KURF and Lhoist North America, especially Mark Luxbacher, for providing us access to the underground site and logistical support. The research has been partially supported by NSF grant 0809696. T. Langford acknowledges support under the National Institute for Standards and Technology American Recovery and Reinvestment Act Measurement Science and Engineering Fellowship Program Award 70NANB10H026 through the University of Maryland.


\bibliographystyle{elsarticle-num}
\bibliography{RiseTime}

\end{document}